\documentclass[twocolumn,preprintnumbers,amsmath,amssymb,aps,prl]{revtex4}

\usepackage{latexsym}
\usepackage{dcolumn}
\usepackage{supertabular,multirow}
\usepackage{epsfig}
\usepackage{bm}
\usepackage{subfigure}
\usepackage{color}

\usepackage{amssymb,amsfonts,amsmath}
\usepackage{euscript}
\usepackage{graphics}
\usepackage{setspace}
\usepackage{graphicx}
\usepackage{hyperref}
\usepackage{xcolor}
\usepackage{textgreek}
\usepackage{siunitx}

\definecolor{brown}{rgb}{0.59, 0.29, 0.0}
\definecolor{orange}{RGB}{255,127,0}
\definecolor{brightube}{rgb}{0.82, 0.62, 0.91}
\definecolor{mauve}{rgb}{0.5, 0.0, 0.9}

\newcommand{\deleted}[1]{}

\newcommand{\bs}[1]{\boldsymbol{#1}}
\newcommand{\pare}[1]{\left( #1 \right)}
\newcommand{\avg}[1]{\langle #1 \rangle}

\begin{document}

\title{Relating microswimmer synthesis to hydrodynamic actuation and rheotactic tunability}

\author{Quentin Brosseau$^1$, Florencio Balboa Usabiaga$^2$,
  Enkeleida Lushi$^3$, Yang Wu$^4$, Leif Ristroph$^1$, Jun Zhang$^1$,
  Michael Ward$^4$ and Michael J. Shelley$^{1,2}$}
\affiliation{
  $^1$ Courant Institute, New York University, New York, NY 10012, USA, \\
  $^2$ Center for Computational Biology, Flatiron Institute,
  New York, NY 10010, USA \\
  $^3$ Department of Mathematics, New Jersey Institute of Technology,
  Newark, NJ 07102, USA\\
  $^4$ Department of Chemistry, New York University, New York, NY 10012, USA}

\date{\today}

\begin{abstract}

We explore the behavior of micron-scale autophoretic Janus (Au/Pt)
rods, having various Au/Pt length ratios, swimming near a wall in an
imposed background flow. We find that their ability to robustly orient
and move upstream, i.e. to rheotax, depends strongly on the Au/Pt ratio,
which is easily tunable in synthesis. Numerical simulations of
swimming rods actuated by a surface slip show a similar rheotactic
tunability when varying the location of the surface slip versus
surface drag. Slip location determines whether swimmers are Pushers
(rear-actuated), Pullers (front-actuated), or in between. Our
simulations and modeling show that Pullers rheotax most robustly due
to their larger tilt angle to the wall, which makes them responsive to
flow gradients. Thus, rheotactic response infers the nature of
difficult to measure flow-fields of an active particle, establishes
its dependence on swimmer type, and shows how Janus rods can be tuned
for flow responsiveness. We demonstrate the effectiveness of a simple
geometric sieve for rheotactic ability.
\end{abstract}



\maketitle

Swimming microorganisms must contend with boundaries and obstacles in 
their natural environments \cite{Saintillan2013, Goldstein2016, Blechinger2016}. Microbial
habitats have ample surfaces, and swimmer concentrations near them
promote attachment and biofilms \cite{Rusconi2015, Conrad2018}. Motile
bacteria and spermatozoa accumulate near boundaries, move along them
\cite{Berke2008, Hulme2008}, and self-organize under confinement
\cite{Denissenko2012, Vladescu2014, Lushi2014,
  Wioland2016}. Microswimmers also exhibit {\it rheotaxis},
i.e. the ability to actively reorient and swim against
    an imposed flow \cite{Marcos2012}. Surfaces are key for
rheotactic response: fluid shear near boundaries results in
hydrodynamic interactions which favor swimmer alignment against the
oncoming flow and prevent swimmer displacements across streamlines
\cite{Hill2007, Kaya2012, Costanzo2012, Kantsler2014,
  Figueroa-Morales2015}. Swimmers with different propulsion mechanisms
-- front-actuated like {\it puller} micro-algae, or rear-actuated like
{\it pusher} bacteria -- exhibit associated dipolar flow fields
\cite{Purcell1977, Drescher2010, Drescher2011} which result in
dissimilar collective motions \cite{Dombrowski2004, Saintillan2007,
  Saintillan2012} and behavior near boundaries or in flows
\cite{Spagnolie2012, Zoettl2012, Contino2015, Lushi2017,
  Bianchi2017, Mathijssen2018}.

Recent advances in the manufacture and design of artificial swimmers
have triggered an acute interest in developing
synthetic mimetic systems \cite{Paxton2006, Howse2007, Moran2011,
  Duan2015, Blechinger2016, Moran2017}. Like their biological
counterparts, artificial swimmers can accumulate near boundaries
\cite{Takagi2014, Brown2016}, navigate along them
\cite{Das2015, Liu2016}, be guided by geometric or chemical patterns
\cite{Simmchen2016, Uspal2016, DaviesWykes2017, Tong2018} or external forces \cite{Tierno2008, Torres2018}, and can
display rheotaxis near planar surfaces \cite{Uspal2015, Palacci2015,
  Ren2017}. While models have been developed to study their locomotion
and behavior \cite{Spagnolie2012, Takagi2014, Spagnolie2015a,
  Potomkin2017}, the relevance of the swimmers' actuation mechanism
and the resulting hydrodynamic contributions to their rheotactic
motion remains an open question. In large part this is
    due to the difficulty in directly assessing swimmers' flow-fields,
    particularly near walls, and relating experimental observations to
    our theoretical understanding of swimmer geometry, hydrodynamics
    and type (i.e., pusher or puller).

\begin{figure}[t]

\centerline{\includegraphics[width=0.9\linewidth]{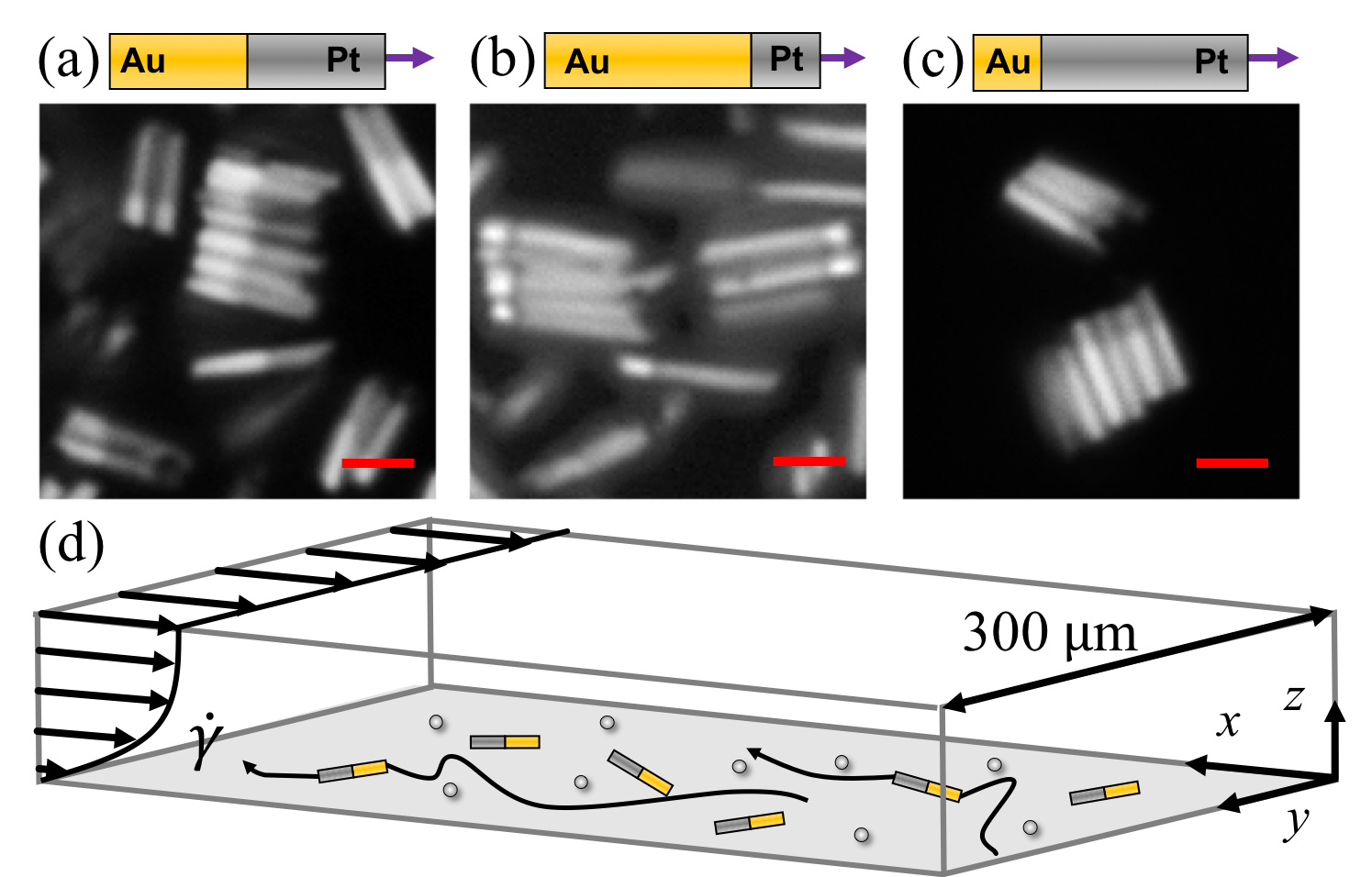}}
 \vspace{-0.15in}
\caption{ 
  The different bimetallic swimmers. The ratio of the metallic
  segments varies from (a) 1:1 for symmetric, to (b) 3:1 for long-gold
  and to (c) 1:3 for long-platinum. Scale bar $1\mu m$. (d) Each
  swimmer type is tested in a rectangular microfluidic channel where
  it is gravitationally confined near the bottom. Under shear flow the
  metallic particles swim upstream.}
\vspace{-0.2in}
\label{fig:rheo_Xp}
\end{figure}

In this Letter, we address this question with
experiments using chemically powered gold-platinum (Au/Pt)
microswimmers combined and compared with numerical simulations.
 In experiments we vary the position of the Au/Pt join along the
 swimmer length, postulating that this varies the location of the flow
 actuation region, and that observed differences in rheotaxis can be
 related to having different pusher- or puller-like swimmers. In
 simulation, we study the rheotactic responses of rod-like
 microswimmers that move through an active surface slip. Different
 placements of the slip region allow us to create pullers, symmetric,
 and pusher microswimmers. We find measurably different rheotactic
 responses in simulation which show quantitative agreement with our
 experiments with Au/Pt active particles conducted in microfluidic
 channels. Lastly, we show that mixed swimmer populations can be
 sorted through a microfluidic sieve that exploits the swimmers'
 different rheotactic behaviors.

{\bf Experimental setup and measurements.} Our Janus microswimmers are
elongated Au/Pt rods, $\sim 2 \mu m$ in length and $\sim 0.3 \mu m$ in
diameter, which propel themselves through self-electrophoresis in
aqueous H$_2$O$_2$ solutions \cite{Moran2011,Moran2017}. The swimmers
are synthesized by electrodeposition \cite{Paxton2006,Banholtzer2009}
to a prescribed ratio of the two metallic segments: symmetric with
Au:Pt (1:1), asymmetric long-gold with Au:Pt (3:1) and asymmetric
long-platinum with Au:Pt (1:3); see Fig. \ref{fig:rheo_Xp}a-c,
details in \cite{SuppMat}.

The swimmers' rheotactic abilities are tested in a rectangular PDMS
microfluidic channel of width $W=300 \mu m$ built following classical
soft-lithography techniques \cite{Qin2010}. We control the background
unidirectional flow down the channel (the $x$-direction) using an
off-stage hydrostatic column. Suspended glass beads of radius
$r_{b}\sim 2.5 \mu m$ serve as markers to measure the flow profile
close to the bottom of the channel where the rods move. We record the
trajectories of swimmers and beads over 1 minute and extract the
instantaneous velocities of swimmers $V_x$ and of beads $U_b$, along
the $x$-axis.
See Fig. \ref{fig:rheo_Xp}d and videos in \cite{SuppMat}.

Thermal fluctuations are important at this scale and the swimmers mean
square displacement for $U_0=0$ at a fixed H$_2$O$_2$ concentration
are used to estimate their translational and rotational diffusivities,
$D_t$ and $D_r$, and deterministic base-line swimming speeds $V_{0}$
\cite{Howse2007,SuppMat}. At fixed H$_2$O$_2$ concentration, swimming
speeds are smaller for asymmetric rods than for symmetric ones,
therefore H$_2$O$_2$ concentration is adjusted to maintain a
comparable velocity $V_0$ between experiments.

The background flow profile close to the wall $U_0(z)$ is measured by
the drift velocity $U_b$ of the suspended glass beads. As the beads
move close to the wall, we find it important to account for the
lubrication forces that act upon them \cite{Goldman1967}. The flow
velocity is estimated to be $U_0(r_b+h_{th})\sim 2.5U_b$ for a thermal
height $h_{th}\sim30nm$\cite{SuppMat}.

{\bf Model and Simulations.} Resolving the chemical and
    electro-hydrodynamics near a wall is challenging. The
    electro-osmotic flow near an self-diffusiophoretic swimmer is the
    result of charge gradients localized on a small surface region
    near the junction of the two metallic segments
    \cite{Paxton2006}. We make the simplifying assumption that this
    results in a surface slip velocity yielding the rod propulsion
    with the Pt segment leading.  As we do not know the extent of the
    slip region, we simply assume that it covers half the rod
    length. The propulsion speed depends on the slip coverage.

We model the swimmer as a rigid, axisymmetric rod immersed in a Stokes
flow and sedimented near an infinite substrate. The rod is discretized
using $N_b$ ``blobs'' at positions $(\bs{r}_i - \bs{q})$ with respect
to the rod center $\bs{q}$ \cite{Delong2015b,Usabiaga2016}. Linear and
angular velocities $\bs{u}$ and $\bs{\omega}$ satisfy the linear
system Eqs. (\ref{eq:balances},\ref{eq:slipCondition}) where
$\bs{\lambda}_i$ are unknown constraint forces enforcing rigid body
motion and $\bs{M}$ is the Rotne-Prager mobility tensor
\cite{Rotne1969} corrected to include the hydrodynamic effect of the
substrate \cite{Blake1971,Swan2007}.

\begin{figure}
\centerline{\includegraphics[width=0.9\linewidth]{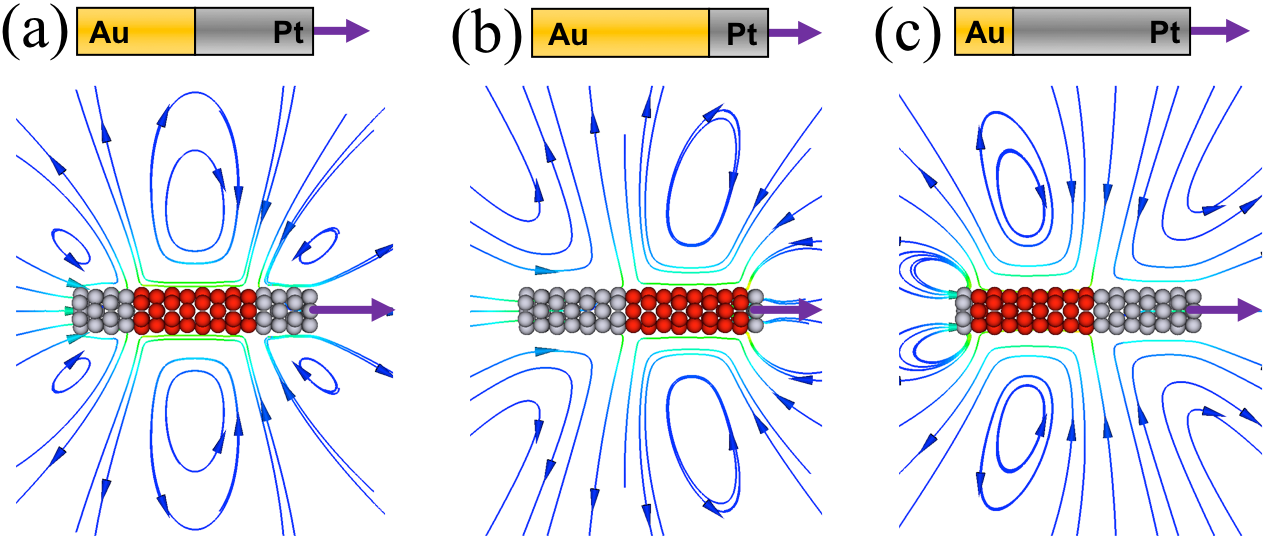}}
 \vspace{-0.1in}
\caption{Computed velocity fields around simulated self-propelled rods
  with a surface slip region (shown in red) (a) at the center, (b) at
  the front, and (c) at the rear, corresponding to, symmetric, puller
  and pusher swimmers, respectively.}
 \vspace{-0.1in}
\label{fig:rheo_Num_1}
\end{figure}

Eq. (\ref{eq:balances}) represents the balance of the geometric
constraint forces with the external force $\bs{F}$ and torque
$\bs{\tau}$ generated by steric interactions with the substrate and
gravity. Eq. (\ref{eq:slipCondition}) gives the balance of fluid,
propulsive, and thermal forces, with $\widetilde{\bs{u}}_i$ the active
slip velocity, $\bs{u}_0(\bs{r}_i)$ the background flow velocity, and
$\sqrt{2k_B T /\Delta t} (\bs{M}^{1/2}\bs{W})_i$ the Brownian noise,
with $k_B$ the Boltzmann constant, $T$ the temperature, $\Delta t$ the
time step, $\bs{W}$ a vector of white noises, and $\bs{M}^{1/2}$
representing the \emph{square root} of the mobility tensor
\cite{Ando2012}. Half the blobs along the rod are ``passive'' with
$\widetilde{\bs{u}}_i=0$, while the other half have an active slip of
constant magnitude $|\widetilde{\bs{u}}_i|=u_s$ parallel to the rod's
main axis. We can set the active slip at the rear, middle, or front;
See Fig. \ref{fig:rheo_Num_1}a-c. After solving
Eqs.~(\ref{eq:balances},\ref{eq:slipCondition}), we update the
configuration with a stochastic integrator \cite{Sprinkle2017}.  Here,
the background flow is linear shear: $\bs{u}_0({\bf
  x})=\dot{\gamma}z\hat{\bf{x}}$.
\begin{align}
&\sum_{i \in (1, N_b)} \bs{\lambda}_i = \bs{F}, \quad \quad
  \sum_{i \in (1, N_b)} (\bs{r}_i - \bs{q}) \times \bs{\lambda}_i = \bs{\tau},
  \label{eq:balances}
\\
&\sum_{j \in (1, N_b)} \bs{M}_{ij} \bs{\lambda}_j = \bs{u} +  \bs{\omega} \times (\bs{r}_i - \bs{q}) - \bs{u}_0(\bs{r}_i) + \widetilde{\bs{u}}_i +
\label{eq:slipCondition}
\\
\nonumber
&\sqrt{2k_B T /\Delta t} \pare{\bs{M}^{1/2} \bs{W}}_i \; \mbox{for } i \in (1, N_b).
\end{align}

Fig. \ref{fig:rheo_Num_1}b \& c show that asymmetrically placed slip
results in a contractile (or puller) dipolar flow for front-slip
particles, and an extensile (or pusher) dipolar flow for rear-slip
particles. The former corresponds to physical long-gold particles, and
the latter to short-gold. Placing the slip region in the middle
(symmetric swimmers) -- see Fig. \ref{fig:rheo_Num_1}a -- yields a
higher-order Stokes quadrupole flow as its leading order
contribution. This corresponds to a symmetric Au/Pt particle.

{\bf Simulation results} To build intuition, we first explore the
simulations' predictions, which motivate a yet simpler dynamical model
of rheotactic response.

\begin{figure}
\centerline{\includegraphics[width=0.9\linewidth]{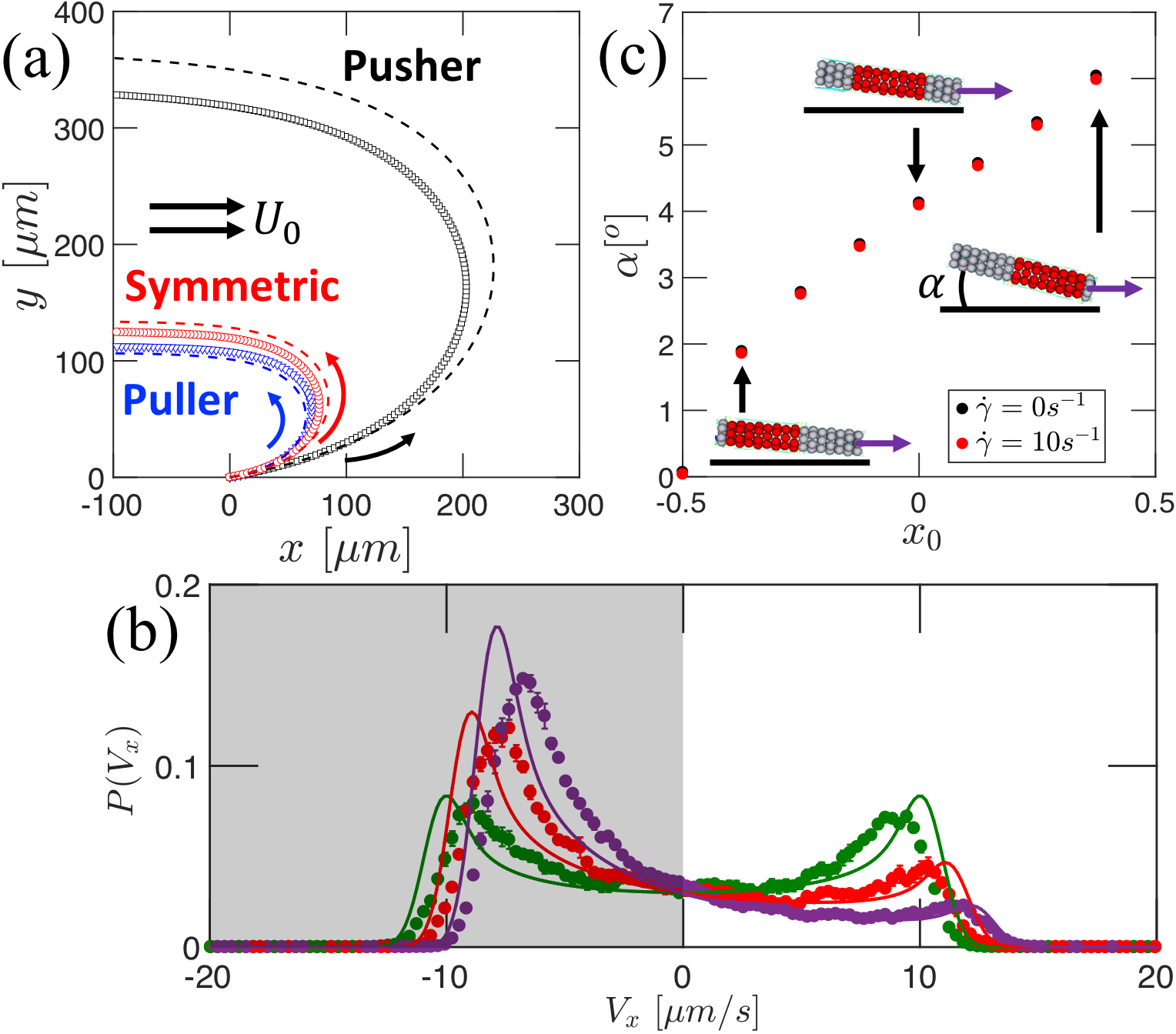}}
\vspace{-0.1in}
\caption{(a) Trajectories of deterministic swimmers with initial
  orientation $\theta_0=\pi/16$, seen from above for simulations
  (solid lines) and theory (dashed lines).
  (b) Particle velocity distribution in the flow direction ($V_x$) for      
  hydrodynamic simulations with brownian noise in a shear flow with
  $\dot{\gamma}$=0$s^{-1}$ ($\color{green}\bullet$), 4$s^{-1}$
  ($\color{red}\bullet$) and 8$s^{-1}$ ($\color{mauve}\bullet$), and
  weathervane model (lines).
  (c) From simulations,
  equilibrium tilt angle relative to the substrate $\alpha$ as a
  function of the center of the slip layer $x_0$ (with
  $x_0=+0.5,0,-0.5$ representing front/middle/aft slip).}
\vspace{-0.2in}
\label{fig:rheo_Num_2}
\end{figure}

Fig. \ref{fig:rheo_Num_2}a illustrates the basic rheotactic response
evinced by our microswimmer model for all swimmer types (pusher,
symmetric, puller). Here, Brownian fluctuations are neglected, and all
swimmers are initially set to swim downstream in a linear shear
flow. In reaction to the background shear each swimmer turns to swim
upstream, with the pusher being the least responsive. For symmetric
swimmers, Fig. \ref{fig:rheo_Num_2}b shows the competition between
rheotaxis induced by flow with thermal fluctuations whose effect is to
de-correlate the swimming direction. In the absence of background flow
($\dot{\gamma}=0$), swimmers diffuse isotropically over long
times. This yields a symmetric bimodal distribution $P(V_{x})$ for the
$x$-velocity $V_{x}$. As the shear-rate becomes increasingly positive,
the distribution becomes asymmetric and increasingly biased towards
upstream swimming (negative $V_{x}$). The distribution curves also
shift right, yielding smaller peak upstream velocities and larger peak
downstream velocities.

Simulations show that active rods swim with a downward tilt towards the
substrate, i.e. with their Pt head segment closer to the wall
\cite{Takagi2014,Ren2017}. The tilt angle $\alpha$
depends weakly on the shear rate but is different for puller, pusher
and symmetric swimmers, see Fig. \ref{fig:rheo_Num_2}c. It is this
tilt that allows the microswimmer to respond to the shear flow near
the wall, and is the origin of rheotaxis. 

The fact of a nonzero tilt angle has been explored most thoroughly by
Spagnolie \& Lauga \cite{Spagnolie2012} who, in seeking to understand
capture of active particles by spherical obstacles \cite{Takagi2014},
numerically studied idealized ellipsoidal ``squirmers'' moving near a
spherical surface. For our numerical model we associate the tilt with
the appearance of high (and low) pressure regions between the swimming
rod and the substrate that tilt the swimmer. These regions appear
where surface velocities, both from slip and no-slip regions, are
converging (and diverging). Moving the slip/no-slip boundary moves the
high pressure region, and thus changes the tilt angle (see
 \cite{SuppMat}).

{\bf A weather-vane model} From these observations we
build an intuitive model displaying a behavior akin to that of a
weathervane. Due to its downward tilt, the shear flow imposes a
larger drag on the tail of a swimming rod.  The drag differential
promotes upstream orientation by producing a torque that depends on the tilt angle $\alpha$. The rod's planar position $\bs{x}=(x,y)$ and
orientation angle $\theta$ evolve as:
\begin{eqnarray}
\label{eq:dotX}
      {\bs{\dot{x}}}=V_0{\bs{n}(\theta)}
      +\dot{\gamma}h{\bs{e}_x}+\sqrt{2D_t}{\bs{W}_x},\\
\label{eq:dotTheta}
\dot{\theta}=\dot{\gamma}\sin\alpha\sin\theta
+\sqrt{2D_r} W_{\theta}.
\end{eqnarray}
Eq. \eqref{eq:dotX} describes a swimming rod that moves with intrinsic
speed $V_0$ at an angle $\theta$
[$\bs{n}=(\cos(\theta),\sin(\theta))$], while is advected by a shear
flow with speed $\dot{\gamma}h$ at a characteristic height $h$ along
the $x$-axis.
Eq. \ref{eq:dotTheta} imposes that the rod angle $\theta$ orients
against the shear flow.
The particle's translational and angular diffusion are $D_t$ and
$D_r$. $\bs{W}_x$ and $W_{\theta}$ are uncorrelated white noise
processes.

\begin{figure*}[t]
\centering
\includegraphics[width=1\textwidth]{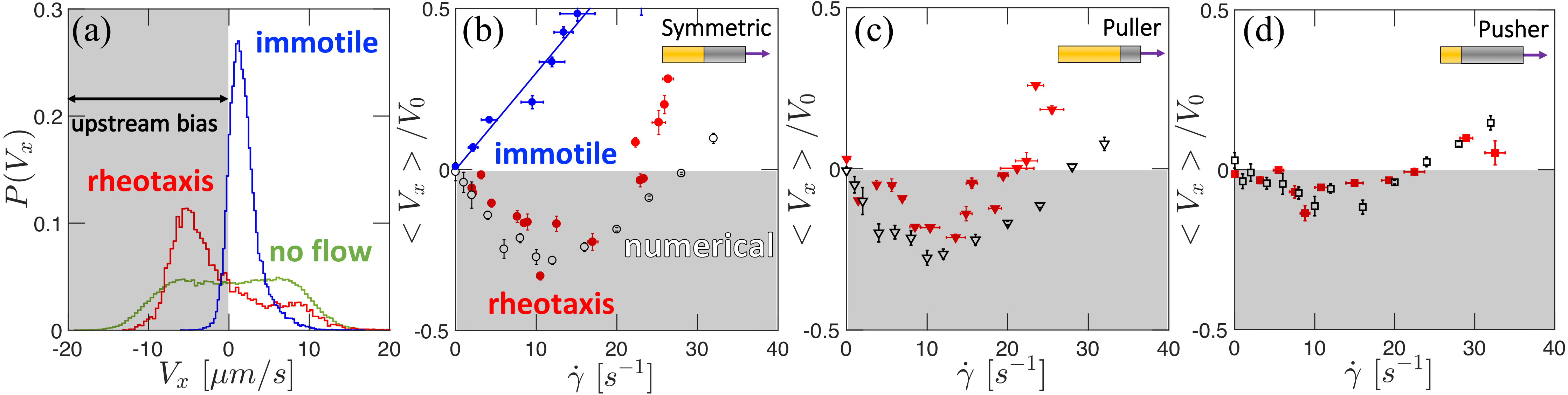}
\vspace{-0.2in}
\caption{(a) From experiments: velocity distribution $P(V_x)$ of
  symmetric swimmers in the absence of background flow
  ($\color{green}-$), with background flow $\dot{\gamma}=8.7 s^{-1}$
  ($\color{red}-$) and for immotile particles in flow
  $\dot{\gamma}=9.5 s^{-1}$ ($\color{blue}-$).  Mean velocity 
{\it vs.} shear rates for (b) symmetric, (c) long-gold puller, and (d)
  long platinum pusher swimmers respectively in experiments
  ($\color{red}+$) and simulations ($\color{black}\circ$) and compared
  to experiment with immotile particles ($\color{blue}+$). Region of
  upstream swimming bias is shaded in gray.}
\vspace{-0.1in}
\label{fig:rheo_results}
\end{figure*}

This model is sufficient to reproduce the deterministic trajectories
of symmetric, puller and pusher swimmers,
Fig.~\ref{fig:rheo_Num_2}a. The tilt angle $\alpha$ controls how fast
a rod reorients against the flow and it explains why pushers are less
responsive to shear flows. The model also predicts a critical swimming
speed to observe positive rheotaxis (upstream swimming).  As
$\dot{\gamma} \rightarrow 0$, the average velocity along the flow is
$\avg{V_x} =\dot{\gamma}\pare{h - V_0 \sin\alpha / 2D_r}$ which sets
the critical speed $V_{0c} = 2D_r h / \sin\alpha$ where the role of
the tilt angle is evident.

From Eqs. \eqref{eq:dotX}-\eqref{eq:dotTheta} we derive the
distribution $P(V_x)$ of the swimmer velocities down the channel
\cite{SuppMat}, see Fig. \ref{fig:rheo_Num_2}b.  Although the
weathervane model neglects hydrodynamics interactions with the
substrate, it agrees with the full numerical simulations for the range
of shear rates and also underlines the influence of parameters
influencing rheotaxis. Note that the absence of lubrication forces
results in an overestimated swimmer velocity in the $x$-direction, a
discrepancy reflected in the distribution peaks shifted to larger
absolute values of $V_x$.


{\bf Experimental validation of the theory.}  In experiments the
velocity distribution $P(V_x)$ follows the same phenomenology
described for the numerical simulations and the reduced model; see
Fig.\ref{fig:rheo_results}a. Under weak shear flow we observe that
passive particles (i.e. no H$_2$O$_2$) are washed downstream whereas
all three types of active rods orient themselves against the flow and
swim upstream.

As suggested by Fig.~\ref{fig:rheo_Num_2}a, both experiments and
simulations reveal that pushers are the least robust
rheotactors. Upstream swimming bias is measured by $\avg{V_{x}}$ as a
function of the shear rate, shown in
Fig. \ref{fig:rheo_results}b-d. Upstream rheotaxis is found for
moderate shear rates, $\dot{\gamma}<20-30\mbox{s}^{-1}$, with the
characteristic non-monotonic trends previously described
\cite{Palacci2015,Ren2017}. The swimmers' ability to move against the
flow reaches a maximum at $\dot{\gamma}\sim 10 \si{s^{-1}}$. When the
viscous drag overcomes the propulsive forces, i.e. $\dot{\gamma}>20
\si{s^{-1}}$, the rods enter a drifting regime characterized by a
rectilinear downstream motion ($\avg{V_x}>0$). For large shear rates
the reduced model predicts a linear average velocity
$\avg{V_x}\sim-V_0 + h\dot{\gamma}$. This trend is consistent with
numerical and experimental results of Fig. \ref{fig:rheo_results}b-d
beyond the minimum of $\avg{V_x}$, though with slightly different
slopes. Even in the drifting regime the average speed
$\avg{V_x}$, of active particles is smaller than immotiles ones
because they are directed and swimming upstream.

Both the symmetric and asymmetric swimmers' rheotactic behavior agrees
with the numerical predictions. This result corroborates the partial
slip model used in the numerical model to describe asymmetric Au/Pt
distributions. Qualitatively, simulations indicate that the maximum
velocity upstream should be larger for puller and symmetric swimmers
than for pushers. Experiments found roughly a factor of two difference
between the maximum upstream velocities between pushers and pullers at
comparable shear values, implying that the reorienting torque is
strongest for pullers. This observation further agrees with the
deterministic trajectories presented in
Fig. \ref{fig:rheo_Num_2}a. There, the parameter that differentiates
those swimmers' dynamics is their tilt angle $\alpha$, identifying it
as a crucial parameter to engineer efficient rheotactors.

\begin{figure}
\centerline{\includegraphics[width=\linewidth]{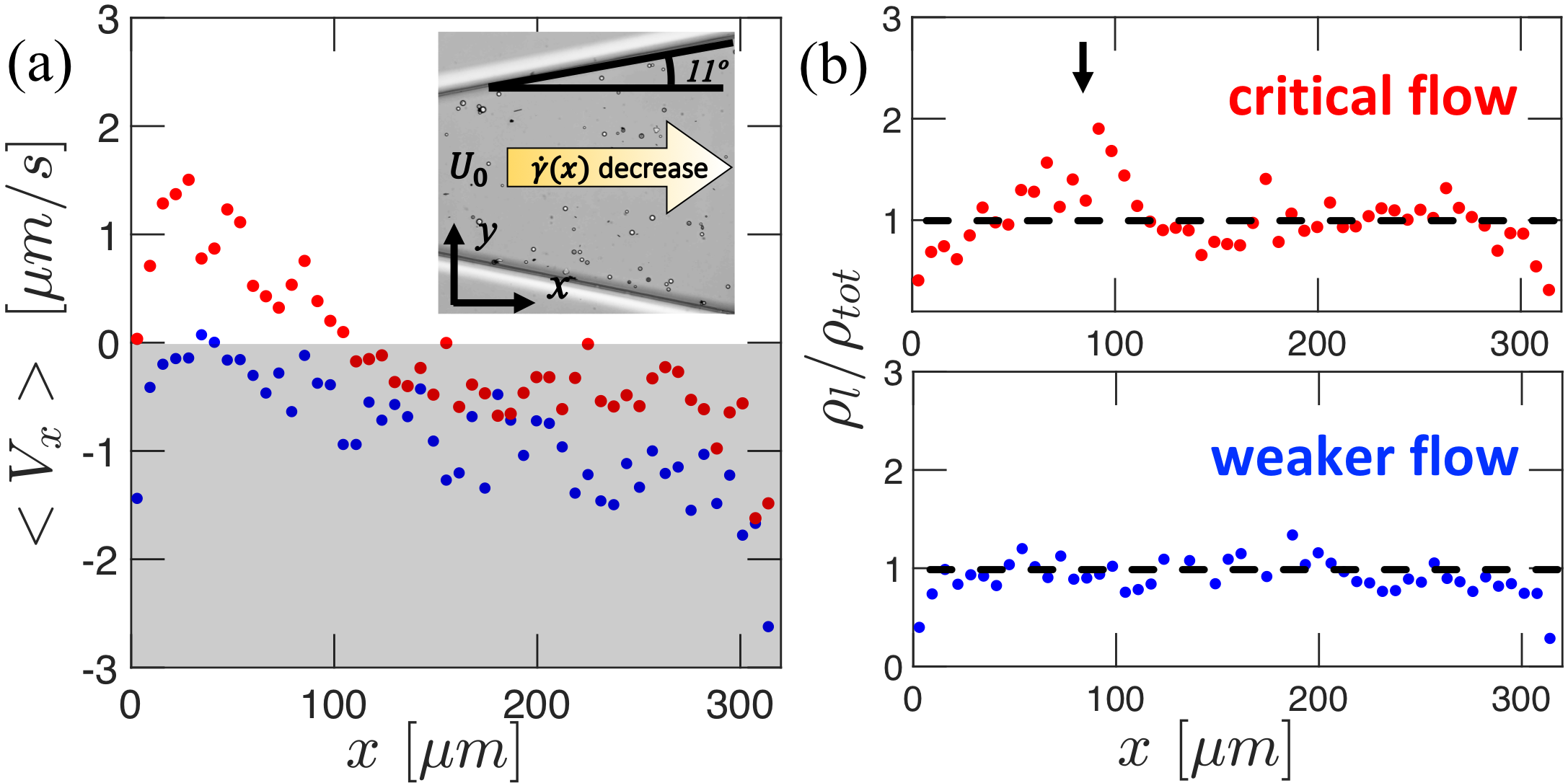}}
\vspace{-0.1in}
\caption{ A rheotactor sieve: (a) The local mean swimmer velocity for
  critical (${\color{red}\bullet}$) and low (${\color{blue}\bullet}$)
  flow rates in a microfluidic sieve geometry (inset). At the critical
  flow rate, particles swim upstream in the wide part of the channel
  and downstream in the narrow part.  (b) The time-integrated swimmer
  density profile normalised by the average swimmer density
  $\rho_l/\rho_{tot}$ reveals a concentration of swimmers where the
  mean swimmer velocity changes sign (indicated by an arrow).}
\vspace{-0.1in}
\label{fig:nozzle}
\end{figure}

{\bf A rheotactor sieve.}  Fig. \ref{fig:nozzle}a presents the concept
of a ``microfluidic sieve'' consisting of a diverging channel. A
constant fluid influx yields a decreasing shear gradient
downstream. We show that in limited windows of flow-rates, the
rheotactors travel upstream in the nozzle until facing a ``shear
wall'' that prevents them from traveling further, thus concentrating
them in those locations.

Fig. \ref{fig:nozzle}b compares the symmetric swimmers' local density
integrated over a period of three minutes for two different shear
regimes. The local swimmer density, $\rho_l$ along the $x$-direction
is normalized by the average density within the channel, $\rho_{tot}$,
to compare experiments with different total numbers of microswimmers.
For small flow-rates (blue), rheotactors swim upstream at any point of
the nozzle. The population density is evenly distributed within the
whole channel as the velocity of the swimmers changes little in this
range of induced shear.  In a critical regime (red), the swimmers
drift downstream from the narrow part of the channel but swim upstream
in the wider sections. The change of sign in the local average swimmer
velocity corresponds roughly with a peak of swimmer density, showing
the accumulation of the swimmers in this region Fig. \ref{fig:nozzle}b
(arrow). This geometry allows the sorting of motile swimmers based on
their speed or tilt angle. From the examples presented above, this
method could conveniently separate mixed populations of asymmetric
swimmers in the same channel.

{\bf Discussion.}  Through experiment, simulation, and modeling, we
demonstrate how to modify rheotactic response by changing swimmer
type, which for Au/Pt Janus rods amounts to changing the location of
the Au/Pt join. Rheotactic tunability is determined primarily by the
tilt angle of the swimmer to the wall, which is controlled by the
distribution of the surface slip.  The quantitative agreement between
experiment and simulation demonstrates that we can infer ``by proxy''
the pusher and puller nature of artificial microswimmers for which
direct flow visualisation is often difficult to obtain. Our study
extends and elaborates upon the recent results of Ren {\it et al.}
\citep{Ren2017} on rheotaxis of symmetric Janus swimmers.

It is chemical reactions that determine the active surface
regions. However, our modeling work here, and that of others
\cite{Spagnolie2012}, show that swimmer-substrate hydrodynamic
interactions are sufficient to produce a tilt angle of the rods and
thus yield rheotaxis. Our conclusions should apply to other swimmer
types besides phoretic particles. A careful treatment of the
electro-chemical reactions could refine the model of the active slip
region used in this work, though solving the electro-chemical
reactions in the presence of thermal fluctuations is far from trivial
\cite{Moran2011}.

The placement of the slip region opens other routes to design
artificial swimmers that have specific interactions with
obstacles. For example, particles that swim with their heads up at a
wall will tend to move away from it \cite{Spagnolie2012}. To explore
this idea we numerically designed swimmers that will tilt up
\cite{SuppMat} by placing an active slip region that covers the nose
back to a point forward of the midpoint. This yields a single high
pressure node in the front half that tilts up the rod. Placing the
slip region on the back half creates a low pressure node on the back
half, yielding the same effect. How to experimentally produce Au/Pt
swimmers with such slip distributions is an interesting question that we are investigating now.

{\bf Acknowledgements.}
This work was supported by NSF-MRSEC program DMR-1420073,
and NSF Grants DMS-1463962 and DMS-1620331.

\bibliography{BiblioK.bib}

\begin{thebibliography}{60}
\expandafter\ifx\csname natexlab\endcsname\relax\def\natexlab#1{#1}\fi
\expandafter\ifx\csname bibnamefont\endcsname\relax
  \def\bibnamefont#1{#1}\fi
\expandafter\ifx\csname bibfnamefont\endcsname\relax
  \def\bibfnamefont#1{#1}\fi
\expandafter\ifx\csname citenamefont\endcsname\relax
  \def\citenamefont#1{#1}\fi
\expandafter\ifx\csname url\endcsname\relax
  \def\url#1{\texttt{#1}}\fi
\expandafter\ifx\csname urlprefix\endcsname\relax\def\urlprefix{URL }\fi
\providecommand{\bibinfo}[2]{#2}
\providecommand{\eprint}[2][]{\url{#2}}

\bibitem[{\citenamefont{Saintillan and Shelley}(2013)}]{Saintillan2013}
\bibinfo{author}{\bibfnamefont{D.}~\bibnamefont{Saintillan}} \bibnamefont{and}
  \bibinfo{author}{\bibfnamefont{M.~E.} \bibnamefont{Shelley}},
  \bibinfo{journal}{Comp. Rend. Phys.}  (\bibinfo{year}{2013}).

\bibitem[{\citenamefont{Goldstein}(2016)}]{Goldstein2016}
\bibinfo{author}{\bibfnamefont{R.~E.} \bibnamefont{Goldstein}},
  \bibinfo{journal}{Journal of Fluid Mechanics} \textbf{\bibinfo{volume}{807}},
  \bibinfo{pages}{1} (\bibinfo{year}{2016}).

\bibitem[{\citenamefont{Blechinger et~al.}(2016)\citenamefont{Blechinger,
  Di~Leonardo, L\"owen, Reichhardt, Volpe, and Volpe}}]{Blechinger2016}
\bibinfo{author}{\bibfnamefont{C.}~\bibnamefont{Blechinger}},
  \bibinfo{author}{\bibfnamefont{R.}~\bibnamefont{Di~Leonardo}},
  \bibinfo{author}{\bibfnamefont{H.}~\bibnamefont{L\"owen}},
  \bibinfo{author}{\bibfnamefont{C.}~\bibnamefont{Reichhardt}},
  \bibinfo{author}{\bibfnamefont{G.}~\bibnamefont{Volpe}}, \bibnamefont{and}
  \bibinfo{author}{\bibfnamefont{G.}~\bibnamefont{Volpe}},
  \bibinfo{journal}{Rev. Mod. Phys.} \textbf{\bibinfo{volume}{88}},
  \bibinfo{pages}{045006} (\bibinfo{year}{2016}).

\bibitem[{\citenamefont{Rusconi and Stocker}(2015)}]{Rusconi2015}
\bibinfo{author}{\bibfnamefont{R.}~\bibnamefont{Rusconi}} \bibnamefont{and}
  \bibinfo{author}{\bibfnamefont{R.}~\bibnamefont{Stocker}},
  \bibinfo{journal}{Curr. Op. Micro.} \textbf{\bibinfo{volume}{25}},
  \bibinfo{pages}{1} (\bibinfo{year}{2015}).

\bibitem[{\citenamefont{Conrad and Poling-Skutvik}(2018)}]{Conrad2018}
\bibinfo{author}{\bibfnamefont{J.}~\bibnamefont{Conrad}} \bibnamefont{and}
  \bibinfo{author}{\bibfnamefont{R.}~\bibnamefont{Poling-Skutvik}},
  \bibinfo{journal}{Annu. Rev. Chem. Biomol. Eng.}
  \textbf{\bibinfo{volume}{9}}, \bibinfo{pages}{175} (\bibinfo{year}{2018}).

\bibitem[{\citenamefont{Berke et~al.}(2008)\citenamefont{Berke, Turner, Berg,
  and Lauga}}]{Berke2008}
\bibinfo{author}{\bibfnamefont{A.}~\bibnamefont{Berke}},
  \bibinfo{author}{\bibfnamefont{L.}~\bibnamefont{Turner}},
  \bibinfo{author}{\bibfnamefont{H.}~\bibnamefont{Berg}}, \bibnamefont{and}
  \bibinfo{author}{\bibfnamefont{E.}~\bibnamefont{Lauga}},
  \bibinfo{journal}{Phys. Rev. Lett.} \textbf{\bibinfo{volume}{101}},
  \bibinfo{pages}{038102} (\bibinfo{year}{2008}).

\bibitem[{\citenamefont{Elizabeth~Hulme
  et~al.}(2008)\citenamefont{Elizabeth~Hulme, DiLuzio, Shevkoplyas, Turner,
  Mayer, Berg, and Whitesides}}]{Hulme2008}
\bibinfo{author}{\bibfnamefont{S.}~\bibnamefont{Elizabeth~Hulme}},
  \bibinfo{author}{\bibfnamefont{W.~R.} \bibnamefont{DiLuzio}},
  \bibinfo{author}{\bibfnamefont{S.~S.} \bibnamefont{Shevkoplyas}},
  \bibinfo{author}{\bibfnamefont{L.}~\bibnamefont{Turner}},
  \bibinfo{author}{\bibfnamefont{M.}~\bibnamefont{Mayer}},
  \bibinfo{author}{\bibfnamefont{H.~C.} \bibnamefont{Berg}}, \bibnamefont{and}
  \bibinfo{author}{\bibfnamefont{G.~M.} \bibnamefont{Whitesides}},
  \bibinfo{journal}{Lab Chip} \textbf{\bibinfo{volume}{8}},
  \bibinfo{pages}{1888} (\bibinfo{year}{2008}).

\bibitem[{\citenamefont{Denissenko et~al.}(2012)\citenamefont{Denissenko,
  Kantsler, Smith, and Kirkman-Brown}}]{Denissenko2012}
\bibinfo{author}{\bibfnamefont{P.}~\bibnamefont{Denissenko}},
  \bibinfo{author}{\bibfnamefont{V.}~\bibnamefont{Kantsler}},
  \bibinfo{author}{\bibfnamefont{D.~J.} \bibnamefont{Smith}}, \bibnamefont{and}
  \bibinfo{author}{\bibfnamefont{J.}~\bibnamefont{Kirkman-Brown}},
  \bibinfo{journal}{Proc. Natl. Acad. Sci.} \textbf{\bibinfo{volume}{109}},
  \bibinfo{pages}{8007} (\bibinfo{year}{2012}).

\bibitem[{\citenamefont{Vladescu et~al.}(2014)\citenamefont{Vladescu, Marsden,
  Schwarz-Linek, Martinez, Arlt, Morozov, Marenduzzo, Cates, and
  Poon}}]{Vladescu2014}
\bibinfo{author}{\bibfnamefont{I.}~\bibnamefont{Vladescu}},
  \bibinfo{author}{\bibfnamefont{E.}~\bibnamefont{Marsden}},
  \bibinfo{author}{\bibfnamefont{J.}~\bibnamefont{Schwarz-Linek}},
  \bibinfo{author}{\bibfnamefont{V.}~\bibnamefont{Martinez}},
  \bibinfo{author}{\bibfnamefont{J.}~\bibnamefont{Arlt}},
  \bibinfo{author}{\bibfnamefont{A.~N.} \bibnamefont{Morozov}},
  \bibinfo{author}{\bibfnamefont{D.}~\bibnamefont{Marenduzzo}},
  \bibinfo{author}{\bibfnamefont{M.}~\bibnamefont{Cates}}, \bibnamefont{and}
  \bibinfo{author}{\bibfnamefont{W.~C.~K.} \bibnamefont{Poon}},
  \bibinfo{journal}{Phys. Rev. Lett} \textbf{\bibinfo{volume}{113}},
  \bibinfo{pages}{268101} (\bibinfo{year}{2014}).

\bibitem[{\citenamefont{Lushi et~al.}(2014)\citenamefont{Lushi, Wioland, and
  Goldstein}}]{Lushi2014}
\bibinfo{author}{\bibfnamefont{E.}~\bibnamefont{Lushi}},
  \bibinfo{author}{\bibfnamefont{H.}~\bibnamefont{Wioland}}, \bibnamefont{and}
  \bibinfo{author}{\bibfnamefont{R.}~\bibnamefont{Goldstein}},
  \bibinfo{journal}{Proc. Nat. Acad. Sci.} \textbf{\bibinfo{volume}{111}},
  \bibinfo{pages}{9733} (\bibinfo{year}{2014}).

\bibitem[{\citenamefont{Wioland et~al.}(2016)\citenamefont{Wioland, Lushi, and
  Goldstein}}]{Wioland2016}
\bibinfo{author}{\bibfnamefont{H.}~\bibnamefont{Wioland}},
  \bibinfo{author}{\bibfnamefont{E.}~\bibnamefont{Lushi}}, \bibnamefont{and}
  \bibinfo{author}{\bibfnamefont{R.}~\bibnamefont{Goldstein}},
  \bibinfo{journal}{New J. Phys.} \textbf{\bibinfo{volume}{18}},
  \bibinfo{pages}{075002} (\bibinfo{year}{2016}).

\bibitem[{\citenamefont{Marcos et~al.}(2012)\citenamefont{Marcos, Fu, Powers,
  and Stocker}}]{Marcos2012}
\bibinfo{author}{\bibfnamefont{M.}~\bibnamefont{Marcos}},
  \bibinfo{author}{\bibfnamefont{H.~C.} \bibnamefont{Fu}},
  \bibinfo{author}{\bibfnamefont{T.~R.} \bibnamefont{Powers}},
  \bibnamefont{and} \bibinfo{author}{\bibfnamefont{R.}~\bibnamefont{Stocker}},
  \bibinfo{journal}{Proc. Nat. Acad. Sci.}  (\bibinfo{year}{2012}), ISSN
  \bibinfo{issn}{0027-8424}.

\bibitem[{\citenamefont{Hill et~al.}(2007)\citenamefont{Hill, Kalkanci,
  McMurry, and Koser}}]{Hill2007}
\bibinfo{author}{\bibfnamefont{J.}~\bibnamefont{Hill}},
  \bibinfo{author}{\bibfnamefont{O.}~\bibnamefont{Kalkanci}},
  \bibinfo{author}{\bibfnamefont{J.~L.} \bibnamefont{McMurry}},
  \bibnamefont{and} \bibinfo{author}{\bibfnamefont{H.}~\bibnamefont{Koser}},
  \bibinfo{journal}{Phys. Rev. Lett.} \textbf{\bibinfo{volume}{98}},
  \bibinfo{pages}{068101} (\bibinfo{year}{2007}).

\bibitem[{\citenamefont{Kaya and Koser}(2012)}]{Kaya2012}
\bibinfo{author}{\bibfnamefont{T.}~\bibnamefont{Kaya}} \bibnamefont{and}
  \bibinfo{author}{\bibfnamefont{H.}~\bibnamefont{Koser}},
  \bibinfo{journal}{Biophys. J.} \textbf{\bibinfo{volume}{102}},
  \bibinfo{pages}{1514 } (\bibinfo{year}{2012}).

\bibitem[{\citenamefont{Costanzo et~al.}(2012)\citenamefont{Costanzo,
  Di~Leonardo, Ruocco, and Angelani}}]{Costanzo2012}
\bibinfo{author}{\bibfnamefont{A.}~\bibnamefont{Costanzo}},
  \bibinfo{author}{\bibfnamefont{R.}~\bibnamefont{Di~Leonardo}},
  \bibinfo{author}{\bibfnamefont{G.}~\bibnamefont{Ruocco}}, \bibnamefont{and}
  \bibinfo{author}{\bibfnamefont{L.}~\bibnamefont{Angelani}},
  \bibinfo{journal}{J. Phys.: Cond. Matt.} \textbf{\bibinfo{volume}{24}},
  \bibinfo{pages}{065101} (\bibinfo{year}{2012}).

\bibitem[{\citenamefont{Kantsler et~al.}(2014)\citenamefont{Kantsler, Dunkel,
  Blayney, and Goldstein}}]{Kantsler2014}
\bibinfo{author}{\bibfnamefont{V.}~\bibnamefont{Kantsler}},
  \bibinfo{author}{\bibfnamefont{J.}~\bibnamefont{Dunkel}},
  \bibinfo{author}{\bibfnamefont{M.}~\bibnamefont{Blayney}}, \bibnamefont{and}
  \bibinfo{author}{\bibfnamefont{R.~E.} \bibnamefont{Goldstein}},
  \bibinfo{journal}{eLife} \textbf{\bibinfo{volume}{111}}
  (\bibinfo{year}{2014}).

\bibitem[{\citenamefont{Figueroa-Morales
  et~al.}(2015)\citenamefont{Figueroa-Morales, Leonardo~Mino, Rivera,
  Caballero, Clement, Altshuler, and Lindner}}]{Figueroa-Morales2015}
\bibinfo{author}{\bibfnamefont{N.}~\bibnamefont{Figueroa-Morales}},
  \bibinfo{author}{\bibfnamefont{G.}~\bibnamefont{Leonardo~Mino}},
  \bibinfo{author}{\bibfnamefont{A.}~\bibnamefont{Rivera}},
  \bibinfo{author}{\bibfnamefont{R.}~\bibnamefont{Caballero}},
  \bibinfo{author}{\bibfnamefont{E.}~\bibnamefont{Clement}},
  \bibinfo{author}{\bibfnamefont{E.}~\bibnamefont{Altshuler}},
  \bibnamefont{and} \bibinfo{author}{\bibfnamefont{A.}~\bibnamefont{Lindner}},
  \bibinfo{journal}{Soft Matt.} \textbf{\bibinfo{volume}{11}},
  \bibinfo{pages}{6284} (\bibinfo{year}{2015}).

\bibitem[{\citenamefont{Purcell}(1977)}]{Purcell1977}
\bibinfo{author}{\bibfnamefont{E.~M.} \bibnamefont{Purcell}},
  \bibinfo{journal}{Am. J. Phys.} \textbf{\bibinfo{volume}{45}},
  \bibinfo{pages}{3} (\bibinfo{year}{1977}).

\bibitem[{\citenamefont{Drescher et~al.}(2011)\citenamefont{Drescher, Dunkel,
  Cisneros, Ganguly, and Goldstein}}]{Drescher2010}
\bibinfo{author}{\bibfnamefont{K.}~\bibnamefont{Drescher}},
  \bibinfo{author}{\bibfnamefont{J.}~\bibnamefont{Dunkel}},
  \bibinfo{author}{\bibfnamefont{L.~H.} \bibnamefont{Cisneros}},
  \bibinfo{author}{\bibfnamefont{S.}~\bibnamefont{Ganguly}}, \bibnamefont{and}
  \bibinfo{author}{\bibfnamefont{R.~E.} \bibnamefont{Goldstein}},
  \bibinfo{journal}{Proc. Nat. Acad. Sci.} \textbf{\bibinfo{volume}{108}},
  \bibinfo{pages}{10940} (\bibinfo{year}{2011}).

\bibitem[{\citenamefont{Drescher et~al.}(2010)\citenamefont{Drescher,
  Goldstein, Michel, Polin, and Tuval}}]{Drescher2011}
\bibinfo{author}{\bibfnamefont{K.}~\bibnamefont{Drescher}},
  \bibinfo{author}{\bibfnamefont{R.~E.} \bibnamefont{Goldstein}},
  \bibinfo{author}{\bibfnamefont{N.}~\bibnamefont{Michel}},
  \bibinfo{author}{\bibfnamefont{M.}~\bibnamefont{Polin}}, \bibnamefont{and}
  \bibinfo{author}{\bibfnamefont{I.}~\bibnamefont{Tuval}},
  \bibinfo{journal}{Phys. Rev. Lett.} \textbf{\bibinfo{volume}{105}},
  \bibinfo{pages}{168101} (\bibinfo{year}{2010}).

\bibitem[{\citenamefont{Dombrowski et~al.}(2004)\citenamefont{Dombrowski,
  Cisneros, Chatkaew, Goldstein, and J.O.}}]{Dombrowski2004}
\bibinfo{author}{\bibfnamefont{C.}~\bibnamefont{Dombrowski}},
  \bibinfo{author}{\bibfnamefont{L.}~\bibnamefont{Cisneros}},
  \bibinfo{author}{\bibfnamefont{S.}~\bibnamefont{Chatkaew}},
  \bibinfo{author}{\bibfnamefont{R.}~\bibnamefont{Goldstein}},
  \bibnamefont{and} \bibinfo{author}{\bibfnamefont{K.}~\bibnamefont{J.O.}},
  \bibinfo{journal}{Phys. Rev. Lett.} \textbf{\bibinfo{volume}{93}},
  \bibinfo{pages}{098103} (\bibinfo{year}{2004}).

\bibitem[{\citenamefont{Saintillan and Shelley}(2007)}]{Saintillan2007}
\bibinfo{author}{\bibfnamefont{D.}~\bibnamefont{Saintillan}} \bibnamefont{and}
  \bibinfo{author}{\bibfnamefont{M.~E.} \bibnamefont{Shelley}},
  \bibinfo{journal}{Phys. Rev. Lett.} \textbf{\bibinfo{volume}{99}},
  \bibinfo{pages}{058102} (\bibinfo{year}{2007}).

\bibitem[{\citenamefont{Saintillan and Shelley}(2012)}]{Saintillan2012}
\bibinfo{author}{\bibfnamefont{D.}~\bibnamefont{Saintillan}} \bibnamefont{and}
  \bibinfo{author}{\bibfnamefont{M.~J.} \bibnamefont{Shelley}},
  \bibinfo{journal}{J. Royal Soc. Int.} \textbf{\bibinfo{volume}{9}},
  \bibinfo{pages}{571} (\bibinfo{year}{2012}).

\bibitem[{\citenamefont{Spagnolie and Lauga}(2012)}]{Spagnolie2012}
\bibinfo{author}{\bibfnamefont{S.~E.} \bibnamefont{Spagnolie}}
  \bibnamefont{and} \bibinfo{author}{\bibfnamefont{E.}~\bibnamefont{Lauga}},
  \bibinfo{journal}{J. Fluid Mech.} \textbf{\bibinfo{volume}{700}},
  \bibinfo{pages}{105} (\bibinfo{year}{2012}).

\bibitem[{\citenamefont{Z\"ottl and Stark}(2012)}]{Zoettl2012}
\bibinfo{author}{\bibfnamefont{A.}~\bibnamefont{Z\"ottl}} \bibnamefont{and}
  \bibinfo{author}{\bibfnamefont{H.}~\bibnamefont{Stark}},
  \bibinfo{journal}{Phys. Rev. Lett.} \textbf{\bibinfo{volume}{108}},
  \bibinfo{pages}{218104} (\bibinfo{year}{2012}).

\bibitem[{\citenamefont{Contino et~al.}(2015)\citenamefont{Contino, Lushi,
  Tuval, Kantsler, and Polin}}]{Contino2015}
\bibinfo{author}{\bibfnamefont{M.}~\bibnamefont{Contino}},
  \bibinfo{author}{\bibfnamefont{E.}~\bibnamefont{Lushi}},
  \bibinfo{author}{\bibfnamefont{I.}~\bibnamefont{Tuval}},
  \bibinfo{author}{\bibfnamefont{V.}~\bibnamefont{Kantsler}}, \bibnamefont{and}
  \bibinfo{author}{\bibfnamefont{M.}~\bibnamefont{Polin}},
  \bibinfo{journal}{Phys. Rev. Lett.} \textbf{\bibinfo{volume}{115}},
  \bibinfo{pages}{258102} (\bibinfo{year}{2015}).

\bibitem[{\citenamefont{Lushi et~al.}(2017)\citenamefont{Lushi, Kantsler, and
  Goldstein}}]{Lushi2017}
\bibinfo{author}{\bibfnamefont{E.}~\bibnamefont{Lushi}},
  \bibinfo{author}{\bibfnamefont{V.}~\bibnamefont{Kantsler}}, \bibnamefont{and}
  \bibinfo{author}{\bibfnamefont{R.~E.} \bibnamefont{Goldstein}},
  \bibinfo{journal}{Phys. Rev. E} \textbf{\bibinfo{volume}{96}},
  \bibinfo{pages}{023102} (\bibinfo{year}{2017}).

\bibitem[{\citenamefont{Bianchi et~al.}(2017)\citenamefont{Bianchi, Saglimbeni,
  and Di~Leonardo}}]{Bianchi2017}
\bibinfo{author}{\bibfnamefont{S.}~\bibnamefont{Bianchi}},
  \bibinfo{author}{\bibfnamefont{F.}~\bibnamefont{Saglimbeni}},
  \bibnamefont{and}
  \bibinfo{author}{\bibfnamefont{R.}~\bibnamefont{Di~Leonardo}},
  \bibinfo{journal}{Phys. Rev. X} \textbf{\bibinfo{volume}{7}},
  \bibinfo{pages}{011010} (\bibinfo{year}{2017}).

\bibitem[{\citenamefont{Mathijssen et~al.}(2018)\citenamefont{Mathijssen,
  Figueroa-Morales, Junot, Clement, Lindner, and Zoettl}}]{Mathijssen2018}
\bibinfo{author}{\bibfnamefont{A.}~\bibnamefont{Mathijssen}},
  \bibinfo{author}{\bibfnamefont{N.}~\bibnamefont{Figueroa-Morales}},
  \bibinfo{author}{\bibfnamefont{G.}~\bibnamefont{Junot}},
  \bibinfo{author}{\bibfnamefont{E.}~\bibnamefont{Clement}},
  \bibinfo{author}{\bibfnamefont{A.}~\bibnamefont{Lindner}}, \bibnamefont{and}
  \bibinfo{author}{\bibfnamefont{A.}~\bibnamefont{Zoettl}},
  \bibinfo{journal}{arXiv} p. \bibinfo{pages}{:1803.01743}
  (\bibinfo{year}{2018}).

\bibitem[{\citenamefont{Paxton et~al.}(2006)\citenamefont{Paxton, Baker, Kline,
  Wang, Mallouk, and Sen}}]{Paxton2006}
\bibinfo{author}{\bibfnamefont{W.~F.} \bibnamefont{Paxton}},
  \bibinfo{author}{\bibfnamefont{P.~T.} \bibnamefont{Baker}},
  \bibinfo{author}{\bibfnamefont{T.~R.} \bibnamefont{Kline}},
  \bibinfo{author}{\bibfnamefont{Y.}~\bibnamefont{Wang}},
  \bibinfo{author}{\bibfnamefont{T.~E.} \bibnamefont{Mallouk}},
  \bibnamefont{and} \bibinfo{author}{\bibfnamefont{A.}~\bibnamefont{Sen}},
  \bibinfo{journal}{J. Am. Chem. Soc.} \textbf{\bibinfo{volume}{128}},
  \bibinfo{pages}{14881} (\bibinfo{year}{2006}).

\bibitem[{\citenamefont{Howse et~al.}(2007)\citenamefont{Howse, Jones, Ryan,
  Gough, Vafabakhsh, and Golestanian}}]{Howse2007}
\bibinfo{author}{\bibfnamefont{J.~R.} \bibnamefont{Howse}},
  \bibinfo{author}{\bibfnamefont{R.~A.~L.} \bibnamefont{Jones}},
  \bibinfo{author}{\bibfnamefont{A.~J.} \bibnamefont{Ryan}},
  \bibinfo{author}{\bibfnamefont{T.}~\bibnamefont{Gough}},
  \bibinfo{author}{\bibfnamefont{R.}~\bibnamefont{Vafabakhsh}},
  \bibnamefont{and}
  \bibinfo{author}{\bibfnamefont{R.}~\bibnamefont{Golestanian}},
  \bibinfo{journal}{Phys. Rev. Lett.} \textbf{\bibinfo{volume}{99}},
  \bibinfo{pages}{048102} (\bibinfo{year}{2007}).

\bibitem[{\citenamefont{Moran and Posner}(2011)}]{Moran2011}
\bibinfo{author}{\bibfnamefont{J.~L.} \bibnamefont{Moran}} \bibnamefont{and}
  \bibinfo{author}{\bibfnamefont{J.~D.} \bibnamefont{Posner}},
  \bibinfo{journal}{J. Fluid Mech.} \textbf{\bibinfo{volume}{680}},
  \bibinfo{pages}{31} (\bibinfo{year}{2011}).

\bibitem[{\citenamefont{Duan et~al.}(2015)\citenamefont{Duan, Wang, Das, Yadav,
  Mallouk, and Sen}}]{Duan2015}
\bibinfo{author}{\bibfnamefont{W.}~\bibnamefont{Duan}},
  \bibinfo{author}{\bibfnamefont{W.}~\bibnamefont{Wang}},
  \bibinfo{author}{\bibfnamefont{S.}~\bibnamefont{Das}},
  \bibinfo{author}{\bibfnamefont{V.}~\bibnamefont{Yadav}},
  \bibinfo{author}{\bibfnamefont{T.~E.} \bibnamefont{Mallouk}},
  \bibnamefont{and} \bibinfo{author}{\bibfnamefont{A.}~\bibnamefont{Sen}},
  \bibinfo{journal}{Ann. Rev. of Ana. Chem.} \textbf{\bibinfo{volume}{8}},
  \bibinfo{pages}{311} (\bibinfo{year}{2015}).

\bibitem[{\citenamefont{Moran and Posner}(2017)}]{Moran2017}
\bibinfo{author}{\bibfnamefont{J.~L.} \bibnamefont{Moran}} \bibnamefont{and}
  \bibinfo{author}{\bibfnamefont{J.~D.} \bibnamefont{Posner}},
  \bibinfo{journal}{Ann. Rev. Fluid Mech.} \textbf{\bibinfo{volume}{49}},
  \bibinfo{pages}{511} (\bibinfo{year}{2017}).

\bibitem[{\citenamefont{Takagi et~al.}(2014)\citenamefont{Takagi, Palacci,
  Braunschweig, Shelley, and Zhang}}]{Takagi2014}
\bibinfo{author}{\bibfnamefont{D.}~\bibnamefont{Takagi}},
  \bibinfo{author}{\bibfnamefont{J.}~\bibnamefont{Palacci}},
  \bibinfo{author}{\bibfnamefont{A.~B.} \bibnamefont{Braunschweig}},
  \bibinfo{author}{\bibfnamefont{M.~J.} \bibnamefont{Shelley}},
  \bibnamefont{and} \bibinfo{author}{\bibfnamefont{J.}~\bibnamefont{Zhang}},
  \bibinfo{journal}{Soft Matt.} \textbf{\bibinfo{volume}{10}},
  \bibinfo{pages}{1784} (\bibinfo{year}{2014}).

\bibitem[{\citenamefont{Brown et~al.}(2016)\citenamefont{Brown, Vladescu,
  Dawson, Vissers, Schwarz-Linek, Lintuvuori, and Poon}}]{Brown2016}
\bibinfo{author}{\bibfnamefont{A.~T.} \bibnamefont{Brown}},
  \bibinfo{author}{\bibfnamefont{I.~D.} \bibnamefont{Vladescu}},
  \bibinfo{author}{\bibfnamefont{A.}~\bibnamefont{Dawson}},
  \bibinfo{author}{\bibfnamefont{T.}~\bibnamefont{Vissers}},
  \bibinfo{author}{\bibfnamefont{J.}~\bibnamefont{Schwarz-Linek}},
  \bibinfo{author}{\bibfnamefont{J.~S.} \bibnamefont{Lintuvuori}},
  \bibnamefont{and} \bibinfo{author}{\bibfnamefont{W.~C.~K.}
  \bibnamefont{Poon}}, \bibinfo{journal}{Soft Matt.}
  \textbf{\bibinfo{volume}{12}}, \bibinfo{pages}{131} (\bibinfo{year}{2016}).

\bibitem[{\citenamefont{Das et~al.}(2015)\citenamefont{Das, Garg, Campbell,
  Howse, Sen, Velegol, Golestanian, and Ebbens}}]{Das2015}
\bibinfo{author}{\bibfnamefont{S.}~\bibnamefont{Das}},
  \bibinfo{author}{\bibfnamefont{A.}~\bibnamefont{Garg}},
  \bibinfo{author}{\bibfnamefont{A.~I.} \bibnamefont{Campbell}},
  \bibinfo{author}{\bibfnamefont{J.}~\bibnamefont{Howse}},
  \bibinfo{author}{\bibfnamefont{A.}~\bibnamefont{Sen}},
  \bibinfo{author}{\bibfnamefont{D.}~\bibnamefont{Velegol}},
  \bibinfo{author}{\bibfnamefont{R.}~\bibnamefont{Golestanian}},
  \bibnamefont{and} \bibinfo{author}{\bibfnamefont{S.~J.}
  \bibnamefont{Ebbens}}, \bibinfo{journal}{Nat. Comm.}
  \textbf{\bibinfo{volume}{6}}, \bibinfo{pages}{8999} (\bibinfo{year}{2015}).

\bibitem[{\citenamefont{Liu et~al.}(2016)\citenamefont{Liu, Zhou, Wang, and
  Zhang}}]{Liu2016}
\bibinfo{author}{\bibfnamefont{C.}~\bibnamefont{Liu}},
  \bibinfo{author}{\bibfnamefont{C.}~\bibnamefont{Zhou}},
  \bibinfo{author}{\bibfnamefont{W.}~\bibnamefont{Wang}}, \bibnamefont{and}
  \bibinfo{author}{\bibfnamefont{H.~P.} \bibnamefont{Zhang}},
  \bibinfo{journal}{Phys. Rev. Lett.} \textbf{\bibinfo{volume}{117}},
  \bibinfo{pages}{198001} (\bibinfo{year}{2016}).

\bibitem[{\citenamefont{Simmchen et~al.}(2016)\citenamefont{Simmchen, Katuri,
  Uspal, Popescu, Tasinkevych, and S{\'a}nchez}}]{Simmchen2016}
\bibinfo{author}{\bibfnamefont{J.}~\bibnamefont{Simmchen}},
  \bibinfo{author}{\bibfnamefont{J.}~\bibnamefont{Katuri}},
  \bibinfo{author}{\bibfnamefont{W.~E.} \bibnamefont{Uspal}},
  \bibinfo{author}{\bibfnamefont{M.~N.} \bibnamefont{Popescu}},
  \bibinfo{author}{\bibfnamefont{M.}~\bibnamefont{Tasinkevych}},
  \bibnamefont{and}
  \bibinfo{author}{\bibfnamefont{S.}~\bibnamefont{S{\'a}nchez}},
  \bibinfo{journal}{Nat. Comm.} \textbf{\bibinfo{volume}{7}},
  \bibinfo{pages}{10598} (\bibinfo{year}{2016}).

\bibitem[{\citenamefont{Uspal et~al.}(2016)\citenamefont{Uspal, Popescu,
  Dietrich, and Tasinkevych}}]{Uspal2016}
\bibinfo{author}{\bibfnamefont{W.~E.} \bibnamefont{Uspal}},
  \bibinfo{author}{\bibfnamefont{M.~N.} \bibnamefont{Popescu}},
  \bibinfo{author}{\bibfnamefont{S.}~\bibnamefont{Dietrich}}, \bibnamefont{and}
  \bibinfo{author}{\bibfnamefont{M.}~\bibnamefont{Tasinkevych}},
  \bibinfo{journal}{Phys. Rev. Lett.} \textbf{\bibinfo{volume}{117}},
  \bibinfo{pages}{048002} (\bibinfo{year}{2016}).

\bibitem[{\citenamefont{Davies-Wykes et~al.}(2017)\citenamefont{Davies-Wykes,
  Zhong, Tong, Adachi, Liu, Ristroph, Ward, Shelley, and
  Zhang}}]{DaviesWykes2017}
\bibinfo{author}{\bibfnamefont{M.~S.} \bibnamefont{Davies-Wykes}},
  \bibinfo{author}{\bibfnamefont{X.}~\bibnamefont{Zhong}},
  \bibinfo{author}{\bibfnamefont{J.}~\bibnamefont{Tong}},
  \bibinfo{author}{\bibfnamefont{T.}~\bibnamefont{Adachi}},
  \bibinfo{author}{\bibfnamefont{Y.}~\bibnamefont{Liu}},
  \bibinfo{author}{\bibfnamefont{L.}~\bibnamefont{Ristroph}},
  \bibinfo{author}{\bibfnamefont{M.~D.} \bibnamefont{Ward}},
  \bibinfo{author}{\bibfnamefont{M.~J.} \bibnamefont{Shelley}},
  \bibnamefont{and} \bibinfo{author}{\bibfnamefont{J.}~\bibnamefont{Zhang}},
  \bibinfo{journal}{Soft Matt.} \textbf{\bibinfo{volume}{13}},
  \bibinfo{pages}{4681} (\bibinfo{year}{2017}).

\bibitem[{\citenamefont{Tong and Shelley}(2018)}]{Tong2018}
\bibinfo{author}{\bibfnamefont{J.}~\bibnamefont{Tong}} \bibnamefont{and}
  \bibinfo{author}{\bibfnamefont{M.}~\bibnamefont{Shelley}},
  \bibinfo{journal}{SIAM Journal on Applied Mathematics}
  \textbf{\bibinfo{volume}{78}}, \bibinfo{pages}{2370} (\bibinfo{year}{2018}).

\bibitem[{\citenamefont{Tierno et~al.}(2008)\citenamefont{Tierno, Golestanian,
  Pagonabarraga, and Sagu\'es}}]{Tierno2008}
\bibinfo{author}{\bibfnamefont{P.}~\bibnamefont{Tierno}},
  \bibinfo{author}{\bibfnamefont{R.}~\bibnamefont{Golestanian}},
  \bibinfo{author}{\bibfnamefont{I.}~\bibnamefont{Pagonabarraga}},
  \bibnamefont{and} \bibinfo{author}{\bibfnamefont{F.}~\bibnamefont{Sagu\'es}},
  \bibinfo{journal}{Phys. Rev. Lett.} \textbf{\bibinfo{volume}{101}},
  \bibinfo{pages}{218304} (\bibinfo{year}{2008}).

\bibitem[{\citenamefont{Garcia-Torres et~al.}(2018)\citenamefont{Garcia-Torres,
  Calero, Sagues, Pagonabarra, and Tierno}}]{Torres2018}
\bibinfo{author}{\bibfnamefont{J.}~\bibnamefont{Garcia-Torres}},
  \bibinfo{author}{\bibfnamefont{C.}~\bibnamefont{Calero}},
  \bibinfo{author}{\bibfnamefont{F.}~\bibnamefont{Sagues}},
  \bibinfo{author}{\bibfnamefont{I.}~\bibnamefont{Pagonabarra}},
  \bibnamefont{and} \bibinfo{author}{\bibfnamefont{P.}~\bibnamefont{Tierno}},
  \bibinfo{journal}{Nat. Comm.} \textbf{\bibinfo{volume}{9}},
  \bibinfo{pages}{491} (\bibinfo{year}{2018}).

\bibitem[{\citenamefont{Uspal et~al.}(2015)\citenamefont{Uspal, Popescu,
  Dietrich, and Tasinkevych}}]{Uspal2015}
\bibinfo{author}{\bibfnamefont{W.}~\bibnamefont{Uspal}},
  \bibinfo{author}{\bibfnamefont{M.~N.} \bibnamefont{Popescu}},
  \bibinfo{author}{\bibfnamefont{S.}~\bibnamefont{Dietrich}}, \bibnamefont{and}
  \bibinfo{author}{\bibfnamefont{M.}~\bibnamefont{Tasinkevych}},
  \bibinfo{journal}{Soft Matter} \textbf{\bibinfo{volume}{11}},
  \bibinfo{pages}{6613} (\bibinfo{year}{2015}).

\bibitem[{\citenamefont{Palacci et~al.}(2015)\citenamefont{Palacci, Sacanna,
  Abramian, Barral, Hanson, Grosberg, Pine, and Chaikin}}]{Palacci2015}
\bibinfo{author}{\bibfnamefont{J.}~\bibnamefont{Palacci}},
  \bibinfo{author}{\bibfnamefont{S.}~\bibnamefont{Sacanna}},
  \bibinfo{author}{\bibfnamefont{A.}~\bibnamefont{Abramian}},
  \bibinfo{author}{\bibfnamefont{J.}~\bibnamefont{Barral}},
  \bibinfo{author}{\bibfnamefont{K.}~\bibnamefont{Hanson}},
  \bibinfo{author}{\bibfnamefont{A.~Y.} \bibnamefont{Grosberg}},
  \bibinfo{author}{\bibfnamefont{D.~J.} \bibnamefont{Pine}}, \bibnamefont{and}
  \bibinfo{author}{\bibfnamefont{P.~M.} \bibnamefont{Chaikin}},
  \bibinfo{journal}{Sci. Adv.} \textbf{\bibinfo{volume}{1}}
  (\bibinfo{year}{2015}).

\bibitem[{\citenamefont{Ren et~al.}(2017)\citenamefont{Ren, Zhou, Mao, Xu,
  Huang, and Mallouk}}]{Ren2017}
\bibinfo{author}{\bibfnamefont{L.}~\bibnamefont{Ren}},
  \bibinfo{author}{\bibfnamefont{D.}~\bibnamefont{Zhou}},
  \bibinfo{author}{\bibfnamefont{Z.}~\bibnamefont{Mao}},
  \bibinfo{author}{\bibfnamefont{P.}~\bibnamefont{Xu}},
  \bibinfo{author}{\bibfnamefont{T.~J.} \bibnamefont{Huang}}, \bibnamefont{and}
  \bibinfo{author}{\bibfnamefont{T.~E.} \bibnamefont{Mallouk}},
  \bibinfo{journal}{ACS Nano} \textbf{\bibinfo{volume}{11}},
  \bibinfo{pages}{10591} (\bibinfo{year}{2017}).

\bibitem[{\citenamefont{Spagnolie et~al.}(2015)\citenamefont{Spagnolie,
  Moreno-Flores, Bartolo, and Lauga}}]{Spagnolie2015a}
\bibinfo{author}{\bibfnamefont{S.~E.} \bibnamefont{Spagnolie}},
  \bibinfo{author}{\bibfnamefont{G.~R.} \bibnamefont{Moreno-Flores}},
  \bibinfo{author}{\bibfnamefont{D.}~\bibnamefont{Bartolo}}, \bibnamefont{and}
  \bibinfo{author}{\bibfnamefont{E.}~\bibnamefont{Lauga}},
  \bibinfo{journal}{Soft Matt.} \textbf{\bibinfo{volume}{11}},
  \bibinfo{pages}{3396} (\bibinfo{year}{2015}).

\bibitem[{\citenamefont{Potomkin et~al.}(2017)\citenamefont{Potomkin, Kaiser,
  Berlyand, and Aranson}}]{Potomkin2017}
\bibinfo{author}{\bibfnamefont{M.}~\bibnamefont{Potomkin}},
  \bibinfo{author}{\bibfnamefont{A.}~\bibnamefont{Kaiser}},
  \bibinfo{author}{\bibfnamefont{L.}~\bibnamefont{Berlyand}}, \bibnamefont{and}
  \bibinfo{author}{\bibfnamefont{I.}~\bibnamefont{Aranson}},
  \bibinfo{journal}{New J. Phys.} \textbf{\bibinfo{volume}{19}},
  \bibinfo{pages}{115005} (\bibinfo{year}{2017}).

\bibitem[{\citenamefont{Banholzer et~al.}(2009)\citenamefont{Banholzer, Qin,
  Millstone, Osberg, and Mirkin}}]{Banholtzer2009}
\bibinfo{author}{\bibfnamefont{M.~J.} \bibnamefont{Banholzer}},
  \bibinfo{author}{\bibfnamefont{L.}~\bibnamefont{Qin}},
  \bibinfo{author}{\bibfnamefont{J.~E.} \bibnamefont{Millstone}},
  \bibinfo{author}{\bibfnamefont{K.~D.} \bibnamefont{Osberg}},
  \bibnamefont{and} \bibinfo{author}{\bibfnamefont{C.~A.}
  \bibnamefont{Mirkin}}, \bibinfo{journal}{Nat. Prot.}
  \textbf{\bibinfo{volume}{4}}, \bibinfo{pages}{838} (\bibinfo{year}{2009}).

\bibitem[{Sup()}]{SuppMat}
\bibinfo{journal}{Supplemental Material} \textbf{\bibinfo{volume}{URL to be
  inserted}} (????).

\bibitem[{\citenamefont{Qin et~al.}(2010)\citenamefont{Qin, Xia, and
  Whitesides}}]{Qin2010}
\bibinfo{author}{\bibfnamefont{D.}~\bibnamefont{Qin}},
  \bibinfo{author}{\bibfnamefont{Y.}~\bibnamefont{Xia}}, \bibnamefont{and}
  \bibinfo{author}{\bibfnamefont{G.~M.} \bibnamefont{Whitesides}},
  \bibinfo{journal}{Nat. Prot.} \textbf{\bibinfo{volume}{5}},
  \bibinfo{pages}{491} (\bibinfo{year}{2010}).

\bibitem[{\citenamefont{Goldman et~al.}(1967)\citenamefont{Goldman, Cox, and
  Brenner}}]{Goldman1967}
\bibinfo{author}{\bibfnamefont{A.}~\bibnamefont{Goldman}},
  \bibinfo{author}{\bibfnamefont{R.}~\bibnamefont{Cox}}, \bibnamefont{and}
  \bibinfo{author}{\bibfnamefont{H.}~\bibnamefont{Brenner}},
  \bibinfo{journal}{Chem. Eng. Sci.} \textbf{\bibinfo{volume}{22}},
  \bibinfo{pages}{637 } (\bibinfo{year}{1967}).

\bibitem[{\citenamefont{Delong et~al.}(2015)\citenamefont{Delong, {Balboa
  Usabiaga}, and Donev}}]{Delong2015b}
\bibinfo{author}{\bibfnamefont{S.}~\bibnamefont{Delong}},
  \bibinfo{author}{\bibfnamefont{F.}~\bibnamefont{{Balboa Usabiaga}}},
  \bibnamefont{and} \bibinfo{author}{\bibfnamefont{A.}~\bibnamefont{Donev}},
  \bibinfo{journal}{J. Chem. Phys.} \textbf{\bibinfo{volume}{143}},
  \bibinfo{pages}{144107} (\bibinfo{year}{2015}).

\bibitem[{\citenamefont{{Balboa Usabiaga} et~al.}(2016)\citenamefont{{Balboa
  Usabiaga}, Kallemov, Delmotte, Bhalla, Griffith, and Donev}}]{Usabiaga2016}
\bibinfo{author}{\bibfnamefont{F.}~\bibnamefont{{Balboa Usabiaga}}},
  \bibinfo{author}{\bibfnamefont{B.}~\bibnamefont{Kallemov}},
  \bibinfo{author}{\bibfnamefont{B.}~\bibnamefont{Delmotte}},
  \bibinfo{author}{\bibfnamefont{A.~P.~S.} \bibnamefont{Bhalla}},
  \bibinfo{author}{\bibfnamefont{B.~E.} \bibnamefont{Griffith}},
  \bibnamefont{and} \bibinfo{author}{\bibfnamefont{A.}~\bibnamefont{Donev}},
  \bibinfo{journal}{Comm. App. Math. Comp. Sci.} \textbf{\bibinfo{volume}{11}},
  \bibinfo{pages}{217} (\bibinfo{year}{2016}).

\bibitem[{\citenamefont{Rotne and Prager}(1969)}]{Rotne1969}
\bibinfo{author}{\bibfnamefont{J.}~\bibnamefont{Rotne}} \bibnamefont{and}
  \bibinfo{author}{\bibfnamefont{S.}~\bibnamefont{Prager}},
  \bibinfo{journal}{J. Chem. Phys.} \textbf{\bibinfo{volume}{50}},
  \bibinfo{pages}{4831} (\bibinfo{year}{1969}).

\bibitem[{\citenamefont{Blake}(1971)}]{Blake1971}
\bibinfo{author}{\bibfnamefont{J.~R.} \bibnamefont{Blake}},
  \bibinfo{journal}{Math. Proc. Cam. Phil. Soc.} \textbf{\bibinfo{volume}{70}},
  \bibinfo{pages}{303} (\bibinfo{year}{1971}).

\bibitem[{\citenamefont{Swan and Brady}(2007)}]{Swan2007}
\bibinfo{author}{\bibfnamefont{J.~W.} \bibnamefont{Swan}} \bibnamefont{and}
  \bibinfo{author}{\bibfnamefont{J.~F.} \bibnamefont{Brady}},
  \bibinfo{journal}{Phys. Fluids} \textbf{\bibinfo{volume}{19}},
  \bibinfo{pages}{113306} (\bibinfo{year}{2007}).

\bibitem[{\citenamefont{Ando et~al.}(2012)\citenamefont{Ando, Chow, Saad, and
  Skolnick}}]{Ando2012}
\bibinfo{author}{\bibfnamefont{T.}~\bibnamefont{Ando}},
  \bibinfo{author}{\bibfnamefont{E.}~\bibnamefont{Chow}},
  \bibinfo{author}{\bibfnamefont{Y.}~\bibnamefont{Saad}}, \bibnamefont{and}
  \bibinfo{author}{\bibfnamefont{J.}~\bibnamefont{Skolnick}},
  \bibinfo{journal}{J. Chem. Phys.} \textbf{\bibinfo{volume}{137}},
  \bibinfo{pages}{064106} (\bibinfo{year}{2012}).

\bibitem[{\citenamefont{Sprinkle et~al.}(2017)\citenamefont{Sprinkle, {Balboa
  Usabiaga}, Patankar, and Donev}}]{Sprinkle2017}
\bibinfo{author}{\bibfnamefont{B.}~\bibnamefont{Sprinkle}},
  \bibinfo{author}{\bibfnamefont{F.}~\bibnamefont{{Balboa Usabiaga}}},
  \bibinfo{author}{\bibfnamefont{N.~A.} \bibnamefont{Patankar}},
  \bibnamefont{and} \bibinfo{author}{\bibfnamefont{A.}~\bibnamefont{Donev}},
  \bibinfo{journal}{J. Chem. Phys.} \textbf{\bibinfo{volume}{147}},
  \bibinfo{pages}{244103} (\bibinfo{year}{2017}).

\end{thebibliography}

\end{document}